\documentclass[12pt]{article} 
\usepackage[sectionbib]{natbib}
\usepackage{array,epsfig,fancyheadings,rotating}
\usepackage[]{hyperref}  
\usepackage{sectsty, secdot}
\sectionfont{\fontsize{12}{14pt plus.8pt minus .6pt}\selectfont}
\renewcommand{\theequation}{\thesection\arabic{equation}}
\subsectionfont{\fontsize{12}{14pt plus.8pt minus .6pt}\selectfont}

\usepackage{amsmath}
\usepackage{amssymb}
\usepackage{amsfonts}
\usepackage{multirow}
\usepackage{amsthm}
\usepackage{amsfonts}
\usepackage{bbm}
\usepackage{wrapfig}
\usepackage{todonotes}
\usepackage{soul}

\newtheorem{theorem}{Theorem}

\theoremstyle{definition}

\usepackage{color}
\begin{document}
	
	
	\renewcommand{\baselinestretch}{2}
	
	\markright{ \hbox{\footnotesize\rm Statistica Sinica
		}\hfill\\[-13pt]
		\hbox{\footnotesize\rm
		}\hfill }
	
	\markboth{\hfill{\footnotesize\rm FIRSTNAME1 LASTNAME1 AND FIRSTNAME2 
	LASTNAME2} \hfill}
	{\hfill {\footnotesize\rm Reference chart for age-varying zero-inflated data} \hfill}
	
	\renewcommand{\thefootnote}{}
	$\ $\par
	
	
	\fontsize{12}{14pt plus.8pt minus .6pt}\selectfont \vspace{0.8pc}
	\centerline{\large\bf ZIKQ: An innovative centile chart method for utilizing  }
	\vspace{2pt} 
	\centerline{\large\bf natural history data in rare disease clinical development}
	\vspace{.4cm} 
	\centerline{Tianying Wang} 
	\vspace{.4cm} 
	\centerline{\it Department of Statistics, Colorado State University, Fort Collins, 
	Colorado, U.S.A.}
	\vspace{.4cm} 
	\centerline{Wenfei Zhang} 
	\vspace{.4cm} 
	\centerline{\it Sarepta Therapeutics, Cambridge, Massachusetts, U.S.A.}
	\vspace{.4cm} 
	\centerline{Ying Wei} 
	\vspace{.4cm} 
	\centerline{\it Department of Biostatistics, Columbia University, New York, New 
	York, U.S.A.}
	
	\vspace{.55cm} \fontsize{9}{11.5pt plus.8pt minus.6pt}\selectfont
	
	
	\begin{quotation}
		\noindent {\it Abstract:}
		Utilizing natural history data as external control  plays an important role in the 
		clinical development of rare diseases, since placebo groups in double-blind 
		randomization trials may not be available due to ethical reasons and low disease 
		prevalence. This article proposed an innovative approach for utilizing natural 
		history data to support  rare disease clinical development by constructing 
		reference centile charts.
		Due to the deterioration nature of certain rare diseases, the distributions of 
		clinical endpoints can be age-dependent and have an absorbing state of zero, 
		which can result in censored natural history data. Existing methods of reference 
		centile charts can not be directly used in the censored natural history data. 
		Therefore, we propose a new calibrated zero-inflated kernel quantile (ZIKQ) 
		estimation to construct reference centile charts from censored natural history 
		data. Using the application to Duchenne Muscular Dystrophy drug 
		development, we demonstrate that the reference centile charts using the ZIKQ 
		method can be implemented to evaluate treatment efficacy and facilitate a 
		more targeted patient enrollment in rare disease clinical development.

		\vspace{9pt}
		\par
	\end{quotation}\par

	\def\thefigure{\arabic{figure}}
	\def\thetable{\arabic{table}}
	
	\renewcommand{\theequation}{\thesection.\arabic{equation}}

	\fontsize{12}{14pt plus.8pt minus .6pt}\selectfont

	\section{Introduction}
	\label{s:intro}
	
	A rare disease is defined as a disease or condition that affects less than 200,000 
	persons in the United States, according to Section 526(a)(2)(A) of the Federal 
	Food, Drug, and Cosmetic Act (FD\&C Act)\citep{ffdca}. There are approximately 
	7,000 recognized rare diseases, cumulatively affecting about 1 in 10 people in the 
	United States. 
	However, most rare diseases do not have approved therapies owing to their 
	complexity and the challenges in clinical development. Most prominently, the 
	golden standard randomization, commonly used in clinical trials 
	\citep{ingram1997case,rubenstein1984effectiveness}, is often unethical and 
	impractical for rare diseases. To determine the treatment efficacy, one has to rely 
	on external controls, which are commonly determined from natural history 
	studies, i.e., preplanned observational studies that ``collect health information in 
	order to understand how a medical condition or disease develops''\citep{nhs}.

\begin{figure}[!ht]
		\centering
		\includegraphics[scale = 0.4]{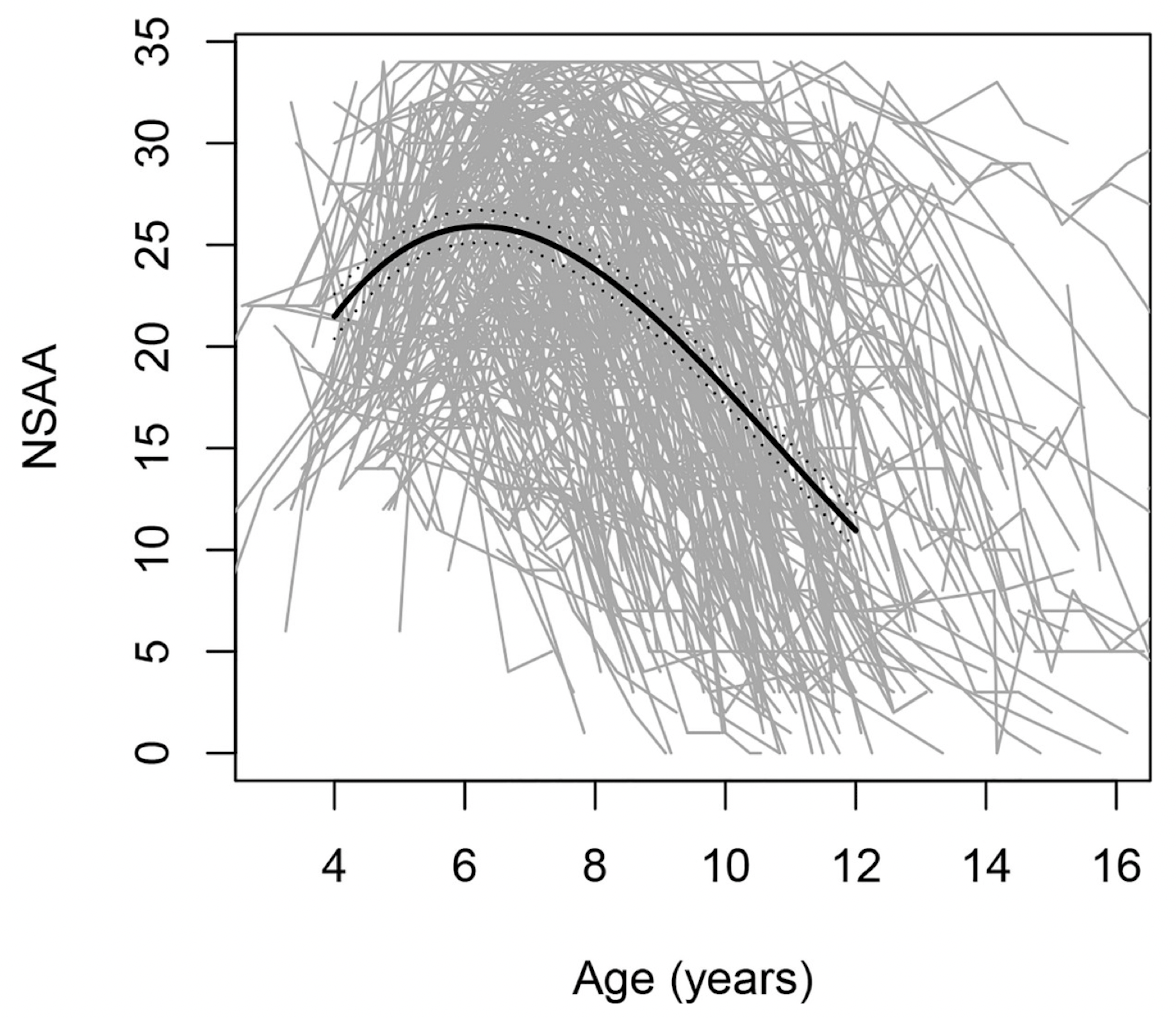}
		\caption{(Source: Figure 1 from \cite{muntoni2019categorising}) NSAA total 
		score trajectories for individual patients by age (in grey) and the fitted mean and 
		95\% confidence interval (in black).}
		\label{fig:early}
	\end{figure}

	By design, natural history data can serve as clinical controls,  as they are 
	non-interventional and often include patients receiving standard of care 
	\citep{ghadessi2020roadmap}.
	For rare diseases with deteriorating conditions, such external/historical control 
	has to be age-dependent to align with the natural disease progression. For 
	example, patients suffering from Duchenne Muscular Dystrophy (DMD), a rare 
	but severe and progressive muscle disorder, typically show a noticeable decline 
	of mobile function by age 3-5 and eventually lose ambulation around 10-12 years 
	old. According to the natural history data of DMD (Figure \ref{fig:early}), the North 
	Star Ambulatory Assessment (NSAA) score, which is a clinical endpoint to 
	evaluate patients' physical functions, tends to increase naturally at an early age 
	4-6,  but later decline gradually until it drops to zero (i.e., "unable to perform 
	independently"). Clearly, the distribution of NSAA is highly age-dependent, with 
	an increased risk of entering the absorbing state of zero.  Therefore using natural 
	history data directly as external control can bring bias due to not-matched age 
	and disease status at baseline. Hence, we propose a new approach to utilize 
	natural history data as external control by constructing the age-adjusted 
	reference centile chart for patients with rare diseases. 
	
	Widely used in pediatrics to monitor children's growth over time, the 
	age-dependent reference centile chart comprises percentile curves of a target 
	measurement over time at selected percentile levels. Typical choices of 
	percentiles are 5\%, 25\%, 50\%, 75\%, and 95\%. 
	When applied to natural history data, such a chart displays the distribution of the 
	target measure in the disease population without interventions.  It helps identify a 
	patient's percentile rank compared to his peers at the same age \textit{without 
	intervention}. Thus, a noticeable increase in a patient's percentile rank indicates a 
	slowdown in disease progression and evidence of the treatment effect.   
	
	Common approaches to building  reference charts include the LMS method 
	\citep{cole1992smoothing} with an age-dependent normality transformation and  
	more general semiparametric quantile-regression approaches proposed in  
	\cite{wei2006quantile,zhang2015regression}. 
	However, those normality-transformation-based methods are designed for 
	continuous variables. They do not incorporate the probability mass on the 
	absorbing state, and hence fail to depict the deteriorating progress of the rare 
	disease. In the case of DMD, a patient physical condition deteriorates over time 
	until he completely loses his ambulatory function, and such a stage is 
	non-reversible. Statistically speaking, the distribution of the clinical endpoint, such 
	as the NSAA score, contains a probability mass on the zero states that increases 
	over time. 
	
	We propose a new calibrated Zero-Inflated Kernel Quantile (ZIKQ) estimation to 
	construct reference centile charts for such deteriorating rare diseases from their 
	natural history data. The resulting reference charts establish 
	theoretically-validated age-dependent external controls for rare diseases, which 
	are essential to evaluate treatment efficacy.  They can also be used to develop 
	enrollment strategies to enhance patient enrollment. Both provide practical 
	solutions to the major challenges in rare disease clinical development --- the 
	scarcity of patients and the difficulty of randomization. 
	
	The rest of the paper is organized as follows. In Section \ref{s:model}, we outline  
	the statistical challenges in estimating reference centiles for rare deteriorating 
	diseases from natural history data,  the proposed methods and algorithms, and 
	how they address the challenges. It is followed by its asymptotic properties 
	discussed in Section \ref{s:asymptotic}. In Section \ref{s:simulation}, we evaluate 
	the performance of our method compared to existing reference chart 
	construction methods. In Section \ref{s:data}, we demonstrate how the estimated 
	reference chart can assist DMD drug development. A discussion is concluded  in 
	Section \ref{s:discuss} with possible future works. Additional simulation results 
	and detailed proofs are provided in the Supplement.

	\section{Methodology}
	\label{s:model}
	
	\subsection{Statistical Model for deteriorating disease progress}
	
	The deteriorating disease progress of a rare disease can be described by a 
	stochastic process $\{Y(t), T_0\}$,  where $Y(t)$ is a continuous non-negative 
	clinical endpoint for a deteriorating disease measured at age
	$t$, and $T_0$ is the age at which $Y$ reaches zero.  The conditional 
	distribution of $Y$ at age t can be decomposed as
	\begin{eqnarray*}
		F(Y\mid t) &=& I\{Y=0\}P(T_0\le t)+F(Y\mid t, T_0>t)P(T_0>t)\\
		&=&\{1-S(t)\}+F(Y\mid t, Y>0)S(t),
	\end{eqnarray*}
	where $S(t) = P(T_0> t)$ is the survival function of $T_0$.
	Consequently, the conditional quantile of $Y$ at age $t$, 
	$Q_Y(\tau\mid t)$, 
	\begin{equation}\label{eqQ}
		Q_Y(\tau\mid t) = \begin{cases}
			0 &\text{$\tau <1-S(t)$} \\
			F^{-1}(\tau^*   \mid t, T_0>t) &\text{$\tau >1-S(t)$},
		\end{cases}
	\end{equation}
	where $ \tau^* = \frac{\tau - \{1-S(t)\}}{S(t)}$. Note that this is a hurdle model, in 
	which the zeros and positive values are clearly separated into two parts. Thus, 
	there is no identifiability issue for the model. One can view $\tau^*$ as a 
	continuous mapping  from $(1-S(t),1)\to(0,1)$. That is, if the target quantile level 
	$\tau > 1-S(t)$, then the quantile function  $ Q_Y(\tau\mid t)$ is equivalent to the 
	conditional quantile function    $F^{-1}(\tau^*   \mid t, T_0>t)$ at the  quantile level 
	$\tau^*$.  That can be derived by solving the equation 
	\begin{eqnarray*}
		\tau 
		&=&\mbox{P}(T_0\le t)+\mbox{P}\{ Y \le Q_Y(\tau\mid t) \mid T_0>t\} 
		\mbox{P}(T_0>t)\\
		&=& \{1-S(t)\}+\mbox{P}\{ Y \le Q_Y(\tau\mid t) \mid T_0>t\} S(t).
	\end{eqnarray*}
	
	Many fatal rare diseases could be caused by rare genetic mutations and 
	chromosome abnormalities other than environmental factors. Owing to the 
	incomplete understanding and the complexity of disease pathophysiology, 
	nonparametric statistical approaches are preferred to model how the clinical 
	endpoint changes with age.
	\subsection{Estimation of $Q_Y(\tau\mid t)$ from a natural history 
	data}\label{ss:estSt}
	A natural history data set consists of $n$ subjects with multiple measurements of 
	interest per subject: $\{(Y_{ij}, t_{ij}): i=1,...,n; j=1,...,J_i\}$, where $t_{ij}$ denotes 
	the observed age for $Y_{ij}$, and $J_i$ is the total number of measurements for 
	the $i$th subject. Such measurement of clinical endpoints often follows a positive 
	monotone trajectory with the increase of age $t$. In reality, patients are very likely 
	to stop visiting hospitals when the disease becomes a severe threat to their 
	physical conditions due to the close to the worst state. To account for such 
	nonignorable missing and impute the barely observed moment of hitting the 
	worst state, i.e., $Y_{ij} = 0$, we follow practical clinical guidance and denote a 
	cutoff $C_0$ close to zero, such that any observations below $C_0$ indicate the 
	future drop to the worst state soon, e.g., after half a year. For simplicity, we 
	assume $Y_{ij} = 0$ if $Y_{i,j-1}<C_0$. In addition to the nonignorable missing, 
	random censoring also happens frequently due to the end of the study, which is a 
	major difference between the natural history data and the traditional 
	time-to-event data. In natural history data, $S(t)$ is modeled as a function of the 
	patient's age but not duration in the study. As the study is often conducted within 
	a certain period and participants joined at different initial ages, they will be 
	censored at different ages when the study ends if the event has not happened 
	yet. This random censoring is independent of the disease progression. Such 
	random censoring does not affect the estimation of $Q_Y(\tau\mid t)$, but needs 
	to be considered in estimating the survival function $S(t)$. 
	
	We consider Kaplan-Meier (KM) estimator \citep{kaplan1958nonparametric} for 
	$S(t)$ and kernel estimation \citep{fan1994robust} for $Q_Y(\tau\mid t)$. KM 
	estimate is a nonparametric maximum likelihood estimate of the survival function, 
	which models the risk as a function of follow-up time.  For brevity, we refer to the 
	clinical endpoint of interest dropping to the worst case as an event or failure 
	happened. The true survival function is $S(t_k) = P(Y_{ij}>0, t_{ij} = t_k)$. Given 
	the hazard function at $t_k$ is $h_k = P(Y_{ij}<C_0|t_{ij} = t_k, Y_{i,j-1}>C_0) = 
	1-\frac{S(t_k)}{S(t_{k-1})}$, the survival function can be written as $S(t_k) = 
	\prod_{r=1}^k (1-h_r)$. Suppose we estimate $S(t)$ on a grid of time points of 
	interest: $\{t_1,\cdots,t_r, \cdots, t_R\}$.
	At a specific $t_r$, we denote  the number of events or failures happened as $d_r 
	= \sum_{i=1}^n \sum_{j=1}^{J_i} \mathbbm{1}(Y_{ij} = 0, t_{ij} = t_r)$, and the 
	individuals randomly dropped out of study as $c_r=\sum_{i=1}^n 
	\mathbbm{1}(Y_{i,J_i}>C_0, t_{r}>t_{i,J_i} > t_{r-1})$. Let $n_r$ be the number of 
	individuals who remained active in the study just before age $t_r$, then $d_r/n_r$ 
	represents the risk of being failed at $t_r$. For the classical KM estimate, all 
	individuals start from the same baseline where the risk of failure is zero and $n_r 
	= n_{r-1} - d_{r-1} - c_{r-1}$, and  $\hat{S}(t_k) = \prod_{r=1}^k 
	\left(1-\frac{d_r}{n_r}\right)$.  
	
	However, directly applying the above procedure is problematic as the life course 
	data observed in natural history studies are different from the time-to-event data 
	studied above. First, the risk in natural history studies is a function of age but not 
	follow-up time. Thus, patients enrolled at different ages have different (unknown) 
	initial risks.  Second, natural history data is a biased sample because the observed 
	data are only representative of people who have survived up to certain age but 
	not those who have reached the worst state and ended their observation before 
	this age. This results in an underestimation when applying $n_r = n_{r-1} - d_{r-1} 
	- c_{r-1}$. A numerical example is provided in Supplement Section S2 to show 
	the bias of the classical design for the KM estimator in the context of 
	time-to-event data.
	
	Based on these two major differences, we propose to use $n_{r,new} = 
	d_{r}+c_{r}+s_r$, where $s_r = \sum_{i=1}^n 
	\sum_{j=1}^{J_i}\mathbbm{1}(Y_{ij}>0, t_{ij}=t_r)$ is the number of individuals 
	remaining active at time $t_r$. Given the two characteristics of the natural history 
	data, a natural plug-in estimator of the hazard function is $\hat h(t_k) = 
	1-\frac{d_r}{n_{r,new}}$ for based on the aforementioned definition of $h_k = 
	P(Y_{ij}<C_0\mid t_{ij} = t_k, Y_{i,j-1}>C_0)$.  The product-limit KM estimator 
	can be constructed naturally with the desired theoretical properties maintained. 
	One can also estimate $S(t)$ from external or historical data or estimate $S(t)$ 
	by other means consistently.

	As for estimating the quantile function $Q_Y(\tau\mid t)$, nonparametric 
	approaches are often preferred in reference chart construction because of their 
	flexibility to capture the nonlinear pattern over age 
	\citep{wei2006quantile,muggeo2013estimating}. We propose to use kernel 
	weighted local linear fitting for nonparametric regression estimation. For any 
	nominal quantile $\tau \in (0,1)$ of $Y$, a characterization of the $\tau$th 
	conditional quantile $Q_Y(\tau\mid t)$ is as
	\begin{eqnarray}\label{eq1}
		Q_Y(\tau\mid t) = Q_Y(\tau^*\mid t, T_0>t) = 
		\arg\min_{a}E\{\rho_{\tau^*}(Y-a)\mid t, T_0>t\},
	\end{eqnarray}
	where $\rho_{\tau^*}(u)$ is the check function given by $\rho_{\tau^*}(u) = 
	u\{\tau^*-I(u<0)\}$.  
	We consider the local linear fitting and approximate $Q_Y(\tau^*\mid t, T_0>t)$ 
	by a linear function:  for $z$ in a neighborhood of $t$, 
	\begin{eqnarray*}Q_Y(\tau^*\mid z, T_0>z) = Q_Y(\tau^*\mid t, 
	T_0>t)+Q'_{Y}(\tau^*\mid t,T_0>t)(z-t)\equiv a_\tau+b_\tau(z-t). 
	\end{eqnarray*}
	Locally, estimating $Q_Y(\tau^*\mid t, T_0>t)$ and $Q'_{Y}(\tau^*\mid t, T_0>t)$ 
	is equivalent to estimating $a_\tau$ and $b_\tau$. Thus, we apply local linear 
	fitting and define the estimator as $\hat Q_Y(\tau\mid t)\equiv \hat a_\tau$, where
	\begin{eqnarray*}
		(\hat a_\tau, \hat b_\tau)= \arg\min_{a,b} \sum_{i,j}
		\mathbbm{1}\{Y_{ij}>0\}\rho_{\tau^*}\left\{Y_{ij} - a-b(t_{ij}-t)\right\} 
		K_{h_{\tau^*}} \left(t_{ij}-t\right),
	\end{eqnarray*}
	and $K(\cdot)$ is a kernel function  with bandwidth $h_{\tau^*}$. Since the above 
	objective function requires the unknown quantity $S(t)$ through ${\tau^*}$, the 
	optimal estimator of $(a_\tau,b_\tau)$ is unattainable. 
	With $\hat{S}(t)$ being a consistent estimator of $S(t)$ and $\hat \tau^* = 
	\frac{\tau - \{1-\hat S(t)\}}{\hat S(t)}$,  the practical estimator $(\tilde{\hat{a}}_\tau, 
	\tilde{\hat{b}}_\tau)$ is
	\begin{eqnarray}\label{esteq}
		(\tilde{\hat{a}}_\tau, \tilde{\hat{b}}_\tau) = \arg\min_{a,b} \sum_{i,j}
		\mathbbm{1}\{Y_{ij}>0\}\rho_{\hat \tau^*}\left\{Y_{ij} - a-b(t_{ij}-t)\right\} K_{ 
		h_{\hat\tau^*}} \left(t_{ij}-t\right),
	\end{eqnarray}
	and the estimated quantile function is 
	\begin{eqnarray*} 
		\hat Q_Y(\tau\mid t) &=& 0\cdot \mathbbm{1}\{\tau \leq 1-\hat 
		S(t)\}+\tilde{\hat{a}}_{\tau}\cdot \mathbbm{1}\{\tau > 1-\hat S(t)\}.
	\end{eqnarray*}

	\subsection{Bandwidth selection}
	We follow the automatic bandwidth selection strategy suggested by 
	\cite{yu1998local} for smoothing conditional quantiles. First, we use the technique 
	of \cite{ruppert1995effective} to select $h_{\rm mean}$. Then, we obtain the 
	bandwidth of $\hat\tau^*$ as 
	\begin{eqnarray*}
		h_{\hat{\tau}^*} = h_{\rm 
		mean}\left[\frac{\hat{\tau}^*(1-\hat{\tau}^*)}{\phi\{\Phi^{-1}(\hat{\tau}^*)\}^2}\right]^{1/5},\end{eqnarray*}
	where $\phi$ and $\Phi$ correspond to the pdf and cdf of standard normal 
	distribution, respectively. Though there are other ready-made approaches to 
	select $h_{\rm mean}$,  additional simulation results suggest that $h_{\rm 
	mean}$ based on the method of \cite{ruppert1995effective} provides  satisfied 
	results across different quantile levels. More details for comparing different 
	bandwidth selection methods are presented in the Supplement.

	As the KM estimator is a step function of time $t$, its smoothness will affect the 
	smoothness of  estimated centile curves. Though we are using the kernel 
	estimation moving through the support of $t$ with carefully selected bandwidth, 
	the discreteness of $\hat{S}(t)$ will be reflected on $\hat{\tau}^*$. One can apply 
	post-smoothing techniques such as B-splines to the estimated chart. However, 
	post-smoothing is not within the scope of this paper, and its resulting properties 
	will not be discussed in detail.

	\section{Asymptotic consistency}\label{s:asymptotic}
	
	In this section, we provide the asymptotic convergence of the chart instead of its 
	asymptotic distribution because of the following two reasons. Practically, 
	reference centile charts are served as a standard criterion once established. Thus, 
	the primary interest is the follow-up investigation but not the inference of the 
	chart itself \citep{wei2006quantile}.  Theoretically, the asymptotic distribution of 
	$Q_Y(\tau\mid t)$ depends on the convergence rate of $\hat S(t)$, while the 
	asymptotic consistency only requires its consistency. To open the possibility of 
	estimating $S(t)$ in either parametric or nonparametric ways, we do not discuss 
	the asymptotic distribution of the chart here. Once the convergence rate of $\hat 
	S(t)$ is given, the asymptotic distribution of $\hat Q_Y(\tau \mid t)$ can be 
	derived based on theoretical proofs in \cite{fan1994robust}. 
	
	For ease of notation, in this section, we simplify the notations and denote 
	observations $\{(Y_{ij}, t_{ij}); i=1,...,n, j = 1,...,J_i\}$ as $\{(Y_{i}, t_i); i=1,...,n\}$ as 
	for the longitudinal information is ignored in the unconditional reference chart. 
	Recall that $\tau^* =  \frac{\tau -\{1-S(t)\}}{S(t)}$ and $\hat\tau^*= \frac{\tau 
	-\{1-\hat S(t)\}}{\hat S(t)}$, we define $\varphi(x\mid t) = 
	E[\rho_{\hat\tau^*}\{Y-m_{\tau^*}(t)+x\mid T=t\}]$, $\varphi'(x\mid t) = \partial 
	\varphi(x\mid t)/\partial t$ and  $\varphi''(x\mid t) = \partial^2 \varphi(x\mid 
	t)/\partial^2 t$. Further, let $f(t) \equiv f_T(t)$ be the density of age $T$ and 
	$g(y\mid t)$ be the conditional density of $Y$ given observed age $T=t$ with 
	respect to measure $\mu$.  Now we state Assumptions (A)-(B) for the kernel 
	estimator and Assumptions (C) for quantile regression as below.  
	
	\paragraph{Assumptions A (for interior points):}
	\begin{enumerate}
		\item[(A1).] The kernel function $K(\cdot)\geq 0$ has a bounded support and  
		satisfies
		$$
		\int K(v)dv = 1, \ \ \int v K(v)dv = 0.
		$$
		\item[(A2).] The density function $f_T(\cdot)$ for $T$ is continuous and $f(t) > 
		0$. 
		\item[(A3).] The function $m_{\tau^*}(\cdot)$ is assumed to have a continuous 
		second derivative. The conditional density function $g(y\mid t)$ is continuous 
		in $t$ for each $y$. 
		\item[(A4).] Assume that there exists positive constants $\epsilon ,\delta$ and a 
		positive function $G(y\mid t)$ such that $\sup_{|t_n-t| \leq \epsilon} g(y\mid 
		t_n)\leq G(y\mid t)$ and that 
		$
		\int | \rho'_{\hat\tau^*}(y-m_{\tau^*}(t))|^{2+\delta}G(y\mid t)d\mu(y)<\infty
		$
		and
		$
		\int \{\rho_{\hat\tau^*}(y-\eta)-\rho_{\hat\tau^*}(y)-\rho'_{\hat\tau^*}(y)\eta 
		\}^2G(y\mid t)d\mu(y)=o(\eta ^2), \ {\rm as\ } \eta \to 0.
		$
		
		\item[(A5).] The quantile check function $\rho_\tau(\cdot)$ is convex with a 
		unique minimizer at $0$.  $\varphi(x\mid z)$, $\varphi'(x\mid z)$ and 
		$\varphi''(x\mid z)$ are functions of $z$ are assumed to be bounded and 
		continuous in a neighborhood of $t$ for all small $x$ and that $\varphi(0\mid 
		t)\neq 0$ for all $t$, including the support boundary, i.e., $t=0$ and $t=1$.
	\end{enumerate}

	\paragraph{Assumptions B (for boundary points):}
	\begin{enumerate}
		\item[(B1).] The kernel function $K(\cdot)\geq 0$ has a bounded support and  
		satisfies
		$$
		\int K(v)dv = 1, \ \ \int v K(v)dv = 0.
		$$
		\item[(B2).] Without loss of generality, we assume the support of the density 
		function $f_T(\cdot)$ is $[0,1]$ and assume $f(0)\equiv \lim_{t \downarrow 0} 
		f_T(t)$ exists and positive.
		\item[(B3).] The function $m_{\tau^*}(\cdot)$ is assumed to have a continuous 
		second derivative. For boundary points $t_n = ch_n$, assume that $g(y\mid 
		0)\equiv \lim_{z\downarrow 0 }g(y\mid z)$ exists. 
		\item[(B4).] There exists positive constants $\epsilon $ and $\delta$ and a 
		positive function $G$ such that $\sup_{t_n \leq \epsilon} g(y\mid t_n)\leq G(y)$ 
		and that 
		$
		\int |\rho'_{\hat\tau^*}(y-m_{\tau^*}(0))|^{2+\delta}G(y)d\mu(y)<\infty
		$
		and
		$
		\int 
		\{\rho_{\hat\tau^*}(y-\eta)-\rho_{\hat\tau^*}(y)-\rho'_{\hat\tau^*}(y)\eta\}^2G(y)d\mu(y)=o(\eta^2),
		 \ {\rm as\ } \eta\to 0,
		$
		where we assume there exists $m_{\tau^*}(0) = \lim_{z\downarrow 0 
		}m_{\tau^*}(z)$.
		\item[(B5).] The quantile check function $\rho_\tau(\cdot)$ is convex with a 
		unique minimizer at $0$.  $\varphi''(x\mid z)$ is a function of $x$ is continuous 
		in a neighborhood of the point $0$, uniformly for $z$ in a neighborhood of $t$.
	\end{enumerate}

	\paragraph{Assumptions C (for quantile regression):}
	\begin{enumerate}
		\item[(C1).] The observations $\{(Y_{i}, t_i); i=1,...,N\}$ can be assumed as i.i.d. 
		from a joint distribution $\mathcal{P}$.
		\item[(C2).] The conditional distribution function $F_Y(\cdot\mid t, t>T_0)$ is 
		absolutely continuous with a positive continuous density $f_{Y\mid 
		t>T_0}(\cdot\mid t)$ on $[0, \infty)$.
		\item[(C3).] The conditional quantile function is right continuous at 0:
		$$
		\lim_{\tau\to \{1-S(t)\}^+} Q_Y(\tau\mid t) =0.
		$$
	\end{enumerate}
	
	Assumptions (A) and (B) are mostly borrowed from 
	\cite{yu1998local,fan1994robust}, with some modifications regarding the 
	calibrated quantile level $\hat{\tau}^*$. Conditions (A1)-(A3) and (B1)-(B3) are 
	necessary for the convergence rate of the bias, and (A4) and (B4) are used for 
	dominated convergence theorem and moment calculation. (A5) and (B5) are 
	satisfied by quantile regression. Thus, the uniqueness of the solution of eq 
	\eqref{esteq} is guaranteed, and the smoothness of the check function 
	$\rho_\tau(\cdot)$ ensures the desirable convergence rate.  Assumptions (C) are 
	similar to \cite{ling2020statistical} and \cite{koenker2005}. Among them, 
	Assumption (C2) ensures the validity of using quantile regression for the positive 
	part, and Assumption (C3) is necessary for the connectivity at the change point. 
	\begin{theorem}
		Under the Assumptions (A), (B) and (C), for any given $\tau\in (0,1)$, 
		$\hat{Q}_Y(\tau\mid t)$ is a consistent estimator, i.e., as $n\to\infty, h_n\to 0$ 
		and $nh_n\to \infty$, 
		$$
		\hat{Q}_Y(\tau\mid t)\xrightarrow{\text{p}} Q_Y(\tau\mid t).
		$$
	\end{theorem}
	
	The asymptotic consistency is constructed separately for both scenarios when 
	$\tau \leq 1-S(t)$ and $\tau > 1-S(t)$ as $Q_Y(\tau \mid  t)$ is defined  
	piecewisely. In particular, when $\tau > 1-S(t)$, we establish the consistency 
	regarding the boundary and interior points for the local linear kernel estimator 
	$\hat Q_Y(\tau\mid t)$.
	The main idea of the proof is similar to the proof in \cite{fan1994robust}, but the 
	loss function is more complicated as $\hat\tau^*$ contains the estimated quantity 
	$\hat S(t)$. Detailed proof is provided in the Supplement.

	\section{Numerical studies}
	\label{s:simulation}
	To evaluate the performance of the proposed kernel quantile regression method 
	for censored growth chart (ZIKQ), we simulate the data mimicking real 
	applications with DMD. 
	
	To mimic the real NSAA score,  the true function $S(t)$ is estimated from Figure 2 
	of \cite{wang2018dmd}, and then the true $\tau$th quantile curve $Q_Y(\tau\mid 
	t)$ is obtained from  Figure 1 of \cite{muntoni2019categorising} based on eq 
	\eqref{eqQ}. The points can be extracted using \textbf{xyscan}, which is a useful 
	tool for extracting points by scanning the plot. We simulate $n = 1,000$ subjects 
	with $J_i$ observations for subject $i$, where $J_i$ is sampled uniformly from 
	the set $\{1, 2, \cdots,6\}$. According to the nature of DMD, we initiate the first 
	observational age of each subject $t_{i1}\sim {\rm Unif} (4, 13)$ and the starting 
	quantile level $\tau_{i1}\sim {\rm Unif} (0, 1)$. Assume patients go to hospitals 
	every six months on a regular basis. The consecutive measurements are collected 
	at  age $t_{ij } = t_{i1}+0.5(j-1),$ and the associated quantile $\tau_{ij}$ is 
	generated from ${\rm Unif}(\max(\tau_{i1} - 0.05, 0), \min(\tau_{i1} + 0.05, 1))$ for 
	$j = 1,\cdots, J_i$. Given $\tau_{ij}$ and $t_{ij}$, we generate the response 
	$Y_{ij}$ as below. If $\tau_{ij}\leq 1-S(t_{ij})$, $Y_{ij}=0$ and no further data for 
	the $i$th individual will be collected. Otherwise, $Y_{ij} = 
	\frac{Q_{\min}(t_{ij})(1-\tau_{ij})+ 
	Q_{\max}(t_{ij})[\tau_{ij}-\{1-S(t_{ij})\}]}{S(t_{ij})}$, where $Q_{\min}(t) = Q_Y\{\tau 
	= 1-S(t)\mid t\}$ and $Q_{\max}(t) = Q_Y(\tau = 1\mid t)$ are derived from the 
	equation $\frac{Q_{\max}(t)-Q_Y(\tau\mid t)}{Q_Y(\tau\mid t) - 
	Q_{\min}(t)}=\frac{1-\tau}{\tau - \{1-S(t)\}}$. 
	
	We compare our method ZIKQ to the two methods discussed in 
	\cite{wei2006quantile}: (1) LMS method (denoted as LMS), implemented by  
	\texttt{R}
	package \texttt{gamlss} \citep{gamlss}; (2) a nonparametric quantile regression 
	method with a B-spline representation of the curves (denoted as QR),  
	implemented using \texttt{R} package \texttt{quantreg} \citep{quantreg}. We 
	report the average estimated curves at quantile levels $\tau = \{10\%, 
	20\%,...,90\%\}$ and root mean square error (RMSE) for each method based on 
	1000 Monte Carlo replicates.
	
	\begin{figure}[!ht]
		\centering
		\includegraphics[scale = 0.35]{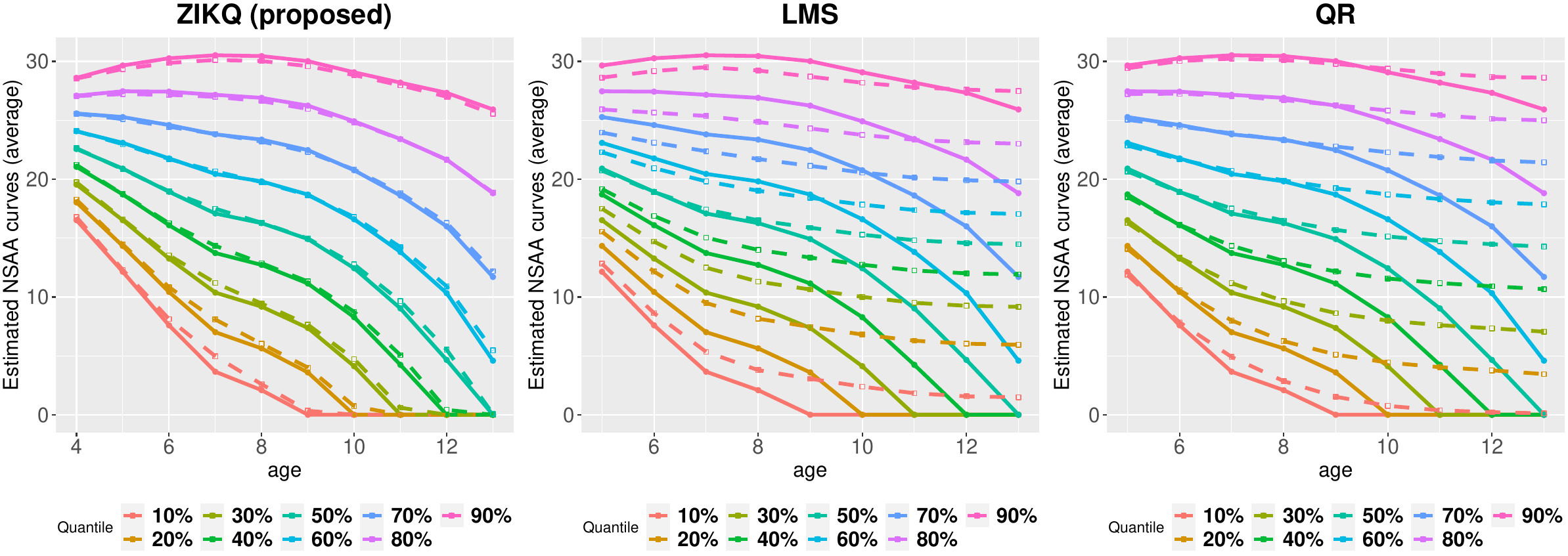}
		\caption{Estimated curves from ZIKQ (left), LMS (middle) and nonparametric 
		quantile regression (right). Solid lines are the ground truth, and dashed lines are 
		averaged results from estimation.}   \label{fig:simulation}
		
	\end{figure}
	
	\begin{table}[!ht]
		\caption{Average RMSE for three methods at different quantile 
		levels.}\label{tab:simu}
		\centering
		\begin{tabular}{rrrrrrrrrr}
			\hline
			Quantile & 10\% & 20\% & 30\% & 40\% & 50\% & 60\% & 70\% & 80\% & 
			90\% \\ 
			\hline
			ZIKQ & 0.44 & 0.57 & 0.66 & 0.76 & 0.81 & 0.87 & 0.78 & 0.68 & 0.57 \\ 
			LMS & 1.78 & 4.13 & 4.89 & 4.77 & 4.02 & 3.14 & 2.40 & 1.87 & 1.11 \\ 
			QR & 0.71 & 2.29 & 3.40 & 4.04 & 4.10 & 3.51 & 2.65 & 1.76 & 0.89 \\ 
			\hline
		\end{tabular}
		
	\end{table}

	Results suggest that both LMS and the quantile regression methods have severe 
	bias, especially after age 8 when over 10\% of individuals are disabled because of 
	disease progression (Figure \ref{fig:simulation}). On the contrary, the proposed 
	ZIKQ method provides consistent estimation with small RMSE for all quantile 
	levels (Table \ref{tab:simu}). We also conducted additional simulations to evaluate 
	multiple choices of bandwidth selection (see Supplement Figure 1). In general, the 
	choice of bandwidth does not significantly affect the estimation results. 
	
	We also conducted additional simulations to evaluate the performance of the 
	proposed ZIKQ method with an irregular observed time grid and under the setting 
	where $Y$ is generated from a stochastic process instead of the quantile 
	functions estimated from natural history data. For both scenarios, the proposed 
	ZIKQ method performs satisfactorily. Detailed results are presented in 
	Supplement Section S2.
	
	\section{Clinical utilities for rare disease treatment developments}
	\label{s:data}
	
	In this section, we use the DMD trial as an example to demonstrate the use of the 
	reference centile chart in rare disease clinical development. That includes (1) 
	understanding the natural course of the disease, (2) assessing treatment efficacy, 
	and (3) informing recruitment and retention strategies. As a  proof of concept, we 
	again use the NSAA score and its smoothed reference centile chart from Section 
	\ref{s:simulation}.  
	
	\subsection{Assessing treatment efficacy}
	Most rare diseases do not have an effective cure. The aim of treatment 
	development is often to slow down the disease progression. By displaying the 
	distributions of clinical endpoints over age, the reference charts provide a 
	comprehensive view of the disease progression under the natural course. 
	For example, according to the natural history data of DMD, the NSAA scores  
	naturally increase at early ages (e.g., 4-7 years old), before they start declining 
	(Figure \ref{fig:early}). The declining rate depends on the patient's age and his 
	initial percentile rank in the population. Therefore,  a change in the NSAA score 
	alone is not sufficient  to determine whether the disease progression was slowed.

	We propose to identify the age-dependent percentile rank of the patients under 
	treatment and view the increased percentile rank as evidence of efficacy. 
	Essentially, the reference chart allows us to compare a patient under treatment to 
	a reference group of  the same age who did not receive interventions (e.g., a 
	pseudo-control group). We hypothesize that a patient without interventions will 
	remain at the same percentile rank in the reference population. Let $q_{i, 0} $ be 
	the percentile rank of the $i$-patient at baseline before the treatment starts. We 
	then measure the
	effectiveness/efficacy of the treatment by 
	the change of one's quantile rank $q_{i,1} - q_{i,0}$, where $q_{i,1}$ is the 
	quantile rank of the $i$th subject at the end of the trial. 
	
\begin{figure}[!ht]
		\centering
		\includegraphics[scale = 0.55]{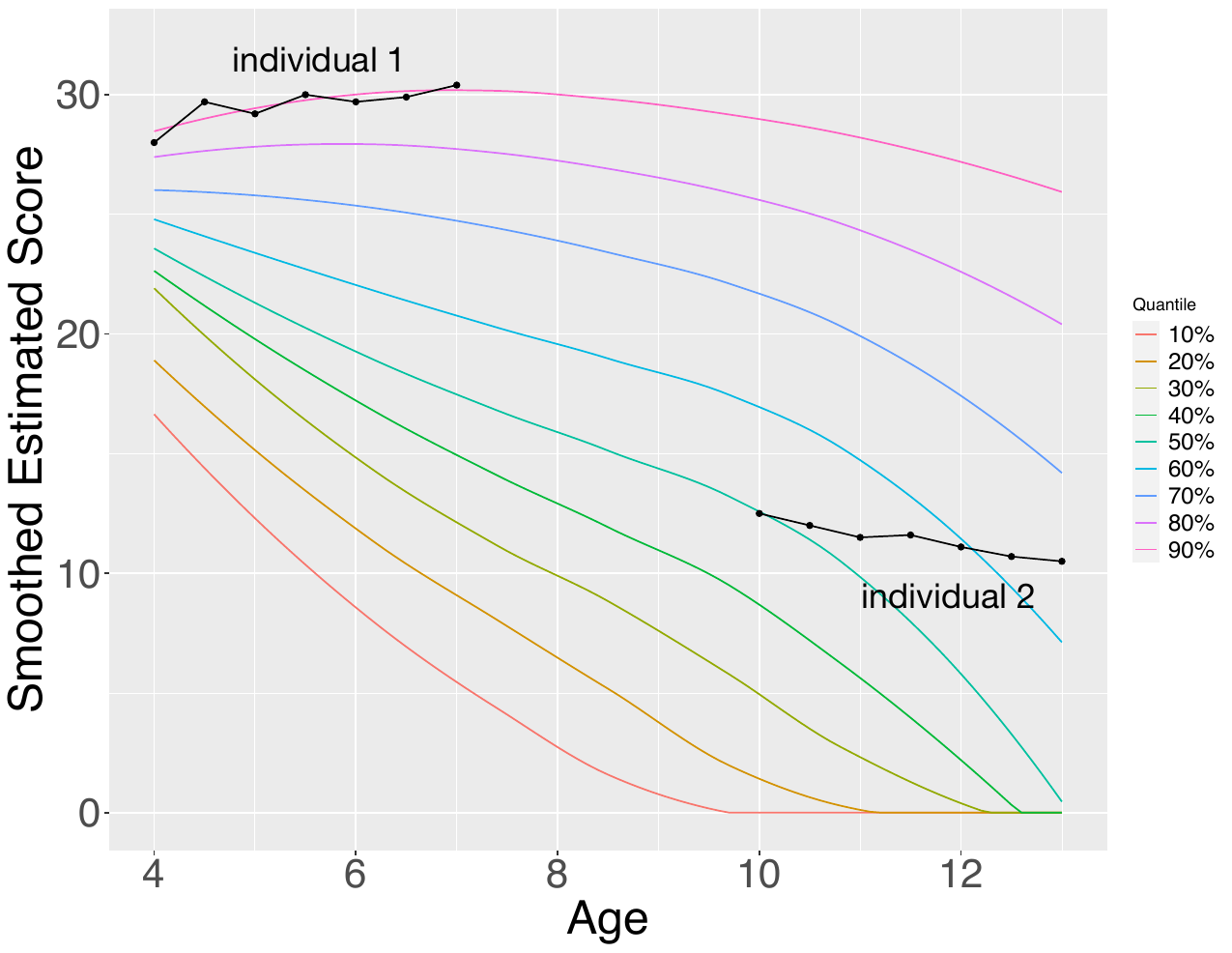}
		\caption{Use reference centile chart to demonstrate treatment effect in early 
		drug development. }
		\label{fig:eg1}
	\end{figure}

	As a demonstration, we plot two trajectories representing two patients under their 
	treatments (Figure \ref{fig:eg1}). Individual 1, whose NSAA score increased but 
	remained on the 90\% quantile curve. We would conclude that the increase in the 
	NSAA score is due to the natural course of the disease at an early age, and there 
	is no evidence of treatment effect. 
	On the contrary, individual 2 experienced a decline in the NSAA score from age 10 
	to 13. However, this decline is much slower than his peers. He started at the 
	50\%  quantile curve(green one), and his percentile rank is above 60th after 
	receiving the treatment. This suggests a potential treatment effect on slowing the 
	disease progression.  
	
	Suppose $n$ patients participated in a  clinical trial on DMD. We can use  
	Wilcoxon signed rank test \citep{wilcoxon1992individual} to determine testing the 
	efficacy.  Let $D_i = q_{i,1} - q_{i,0}$ be the change of the percentile rank of the 
	$i$th patient before and after the treatment.
	We construct a one-side hypothesis test,  $H_0: E (D_i) = 0 \, \ \mbox{v.s.}\   H_a: 
	E (D_i) > 0$. The null hypothesis  $H_0$ suggests no treatment effect, while the 
	alternative hypothesis $H_a$ implies that the treatment slows the disease 
	progression.

	
	After excluding the pairs that $|D_i| = 0$, we order the remaining $N_r$ reduced 
	samples from the smallest absolute differences to the largest absolute 
	differences and obtain their ranks $R_i$. Then, the test statistic $W$ is 
	\begin{eqnarray*}
		W = \sum_{i=1}^{N_r}[sgn(D_i)\cdot R_i].
	\end{eqnarray*}
	With a moderate size of $N_r$, e.g., $N_r > 20$, the $Z$ score can be calculated 
	as $z = W/\sigma_w$, where $\sigma_w = \sqrt{\frac{N_r(N_r+1)(2N_r+1)}{6}}$. 
	Then, the $p$ value can be calculated based on the normal approximation for 
	large samples.

	\subsection{Inform enrollment and retention strategies}
	
\begin{figure}[!ht]
		\centering
		\hspace*{-2cm}
		\includegraphics[scale = 
		0.52]{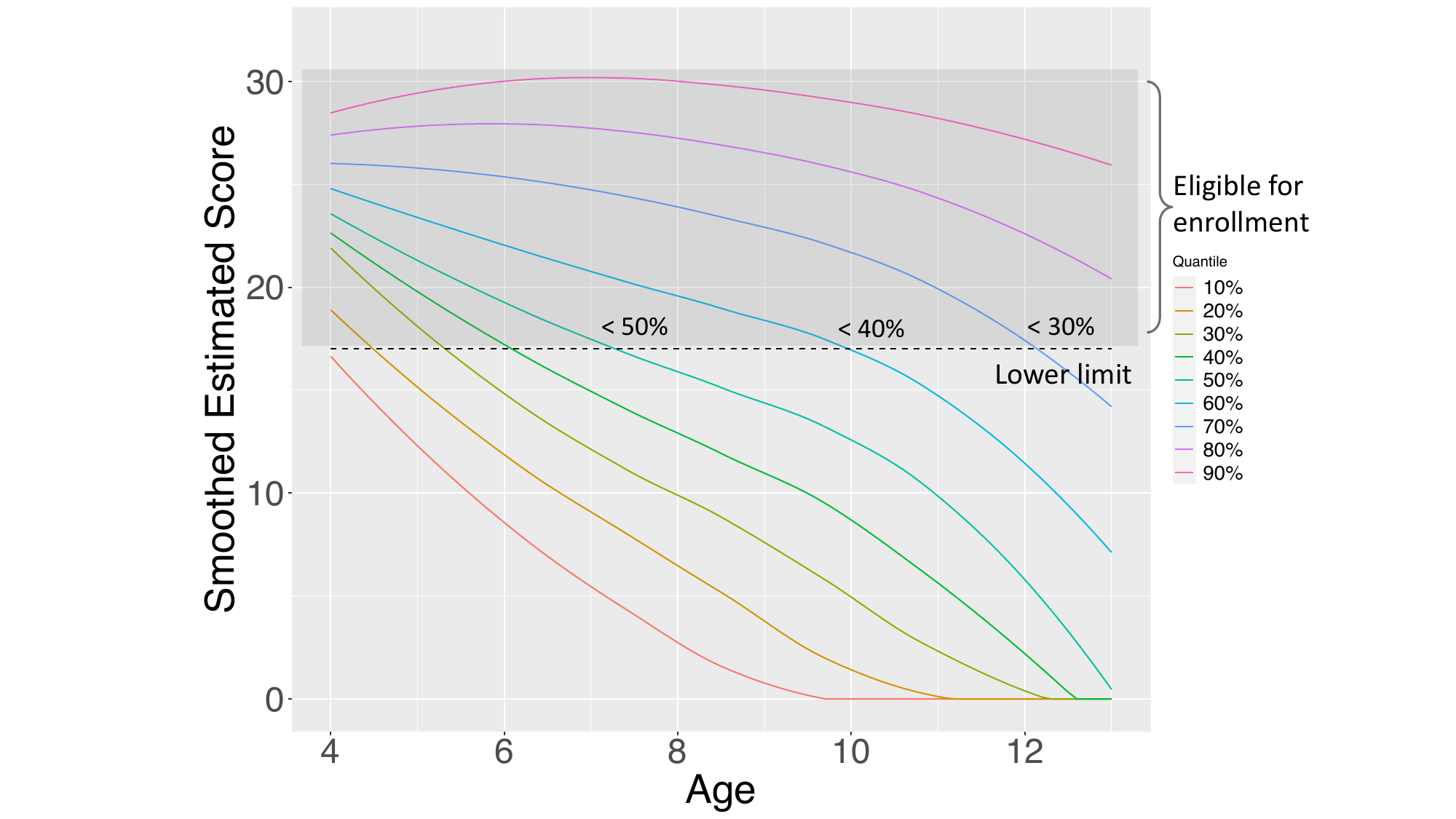}
		\caption{Use reference centile chart to guide patients enrollment criteria. The 
		dashed line indicates NSAA score = 17. The percentages indicated above the 
		line represent the portion of the population qualified for enrollment at the given 
		age.}
		\label{fig:late}
	\end{figure}

	Besides the lack of the capacity for randomization, recruitment and retention are 
	other major challenges of rare disease clinical studies due to the scarcity of 
	patients \citep{crow2018checklist}. A standard inclusion criterion is to
	set a minimum bar to exclude the patients who are too sick to stay on trial.
	Using the DMD trial as an example, the NSAA score of 17 was used as the 
	recruitment lower bound since those patients can stay ambulatory in a two-year 
	study \citep{mazzone201324}. 
	Due to the deteriorating nature of the disease, such fixed recruitment criteria may 
	not be optimal for all age groups. 
	For example, based on the estimated chart (Figure \ref{fig:late}), less than 40\% 
	of patients above ten years old are eligible for enrollment if the NSAA score of 17 
	is used as the minimum inclusion criterion.  That brings the risk of delaying the 
	drug development process due to a lack of effective samples in this age group. 

	Assuming that an individual will maintain the same percentile rank over time 
	without external interference, the reference centile charts could help design an 
	age-dependent enrollment strategy to optimize patient recruitment. Let $q_{i, 
	0}(t) $ be the percentile rank of the $i$th patient at baseline age $t$. We can then 
	predict whether the patient will drop to zero during the trial  by tracing him on the 
	centile chart along his baseline percentile rank. That is, the $i$th patient will be 
	eligible for recruitment if $q_{i, 0}(u) > 0$ for $u\in [t,t+\Delta t]$, where $\Delta 
	t$ is the planned trial duration.

\begin{figure}[!ht]
			\centering
			\includegraphics[scale = 0.45]{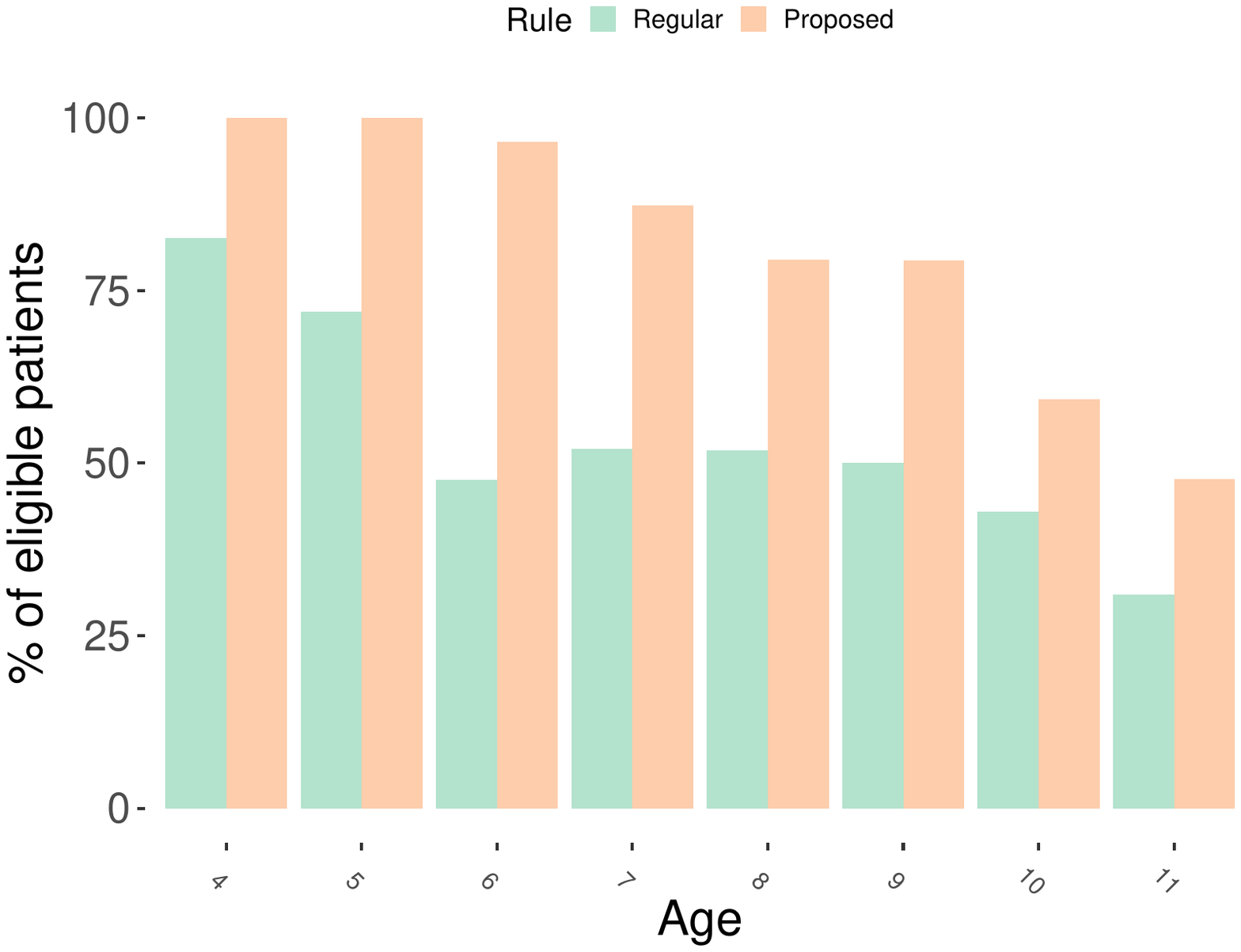}
			\caption{The percentage of eligible patients at different ages by the regular 
			and the proposed inclusion criteria. }
			\label{fig:bar}
		\end{figure}

	Using the same model as the simulation study, we generate a cohort of DMD 
	patients with their  NSAA scores. We then apply the proposed age-dependent 
	recruiting strategy and compare it with the fixed lower bound at NSAA $=17$ 
	(denoted as the "regular rule").   Suppose the study is designed for two years (i.e., 
	$\Delta t=  2$) and the sample size is $n=1000$.  We compare the recruitment 
	results under the regular recruitment rule and our proposed rule regarding the 
	recruitment rate. Using the regular recruitment rule, only  52.9\% of patients are 
	qualified for enrollment, while 81.9\% of patients can be eligible under  
	age-dependent recruitment without undermining the retention rates. Both 
	inclusion  criteria yield a nearly perfect retention rate (100\% vs. 100\%) based on 
	the simulations.  Figure \ref{fig:bar} provides the proportion of eligible patients by 
	age under the two recruitment rules. We observe that age-dependent recruitment 
	could significantly improve the recruitment rate in all age groups, which is crucial 
	to clinical development for rare diseases.


		
		
		
		\section{Discussion}
		\label{s:discuss}
		
		An incomplete understanding of disease pathophysiology is a crucial challenge 
		in the therapeutic development of rare diseases. Over the past decades, 
		researchers have been dedicated to further using comprehensive information 
		from natural history data to help evaluate disease progression and facilitate 
		clinical development. Traditional methods for growth chart construction, such 
		as quantile regression with B-spline and LMS method, have been shown to be 
		heavily biased because of ignoring disease progress and constructing 
		percentiles only based on survived samples. Thus, existing methods cannot be 
		served as a reference for evaluating disease progression. This article provides a 
		powerful tool to construct historical controls from the reference centile chart 
		perspective. Through integrating survival information and adjusting it according 
		to the nature of life-course data, we developed a versatile framework to 
		address the bias issue owing to the nonignorable failure in natural history data. 
		More importantly, we illustrated how the reference chart could benefit clinical 
		development in various ways. Though centile charts are often used without 
		confidence intervals for better interpretability, such as growth chart 
		\citep{wei2006quantile}, one may be interested in constructing confidence 
		intervals. We described how to construct bootstrap-based confidence intervals 
		in Supplement Section S3. Results suggested that the estimation of centile 
		curves is precise as the confidence intervals are narrow and well-separated.

		A practical alternative is to manually impute zeros consecutively after the event 
		occurred so that the sample quantiles will be corrected. That is a naive 
		realization of the proposed method.  After imputation, the techniques applied in 
		the proposed method, i.e., two-part modeling for the zeros and non-zeros, are 
		still required. The imputation-based approach is mathematically equivalent to 
		the survival function correction in the proposed method. However, our 
		approach is more general and rigorous, equipped with asymptotic theory, and 
		can be adjusted by user-specified estimated survival function. This appealing 
		feature allows more accurate estimations when the survival function can be 
		estimated from a large external data set or other historical data. In addition, note 
		that another popular nonparametric analysis approach, B-spline, is not 
		applicable in this framework. As the adjusted quantile level $\tau^*$ is a 
		mapping involving the nominal quantile level $\tau$ and age $t$, simply 
		decomposing the unknown effect of $t$ to $Y$ through B-splines as in 
		\cite{wei2006quantile} will not account for its role in $\tau^*$, while the local 
		fitting can be conducted regarding a fix $\tau$ and $t$. The two competitors, 
		namely LMS and QR, are possibly improved by incorporating the two-stage 
		modeling procedure, especially the first step of estimating $S(t)$. However, 
		ZIKQ could still be more flexible and general, resulting in a more accurate 
		estimation of the centile curves due to the nonparametric nature of kernel 
		estimation.
		
		Given the various practical guidance offered in this article, there is a wide range 
		of future works that are worth investigating. For instance, considering disease 
		development within one subject and making inferences with personal 
		longitudinal information is a promising future direction. Similar to 
		\cite{zhang2015regression}, disease progression can be analyzed based on the 
		proposed reference centile chart functionally to advance our understanding and 
		insights into rare diseases. When there are covariates that need to be adjusted, 
		the current framework can be extended to the realm of censored quantile 
		regression \citep{wang2009locally,portnoy2003censored}.  Built on the 
		KM-type estimator for $S(t)$, one of the key assumptions, which is also the key 
		assumption for all KM-type estimators, is noninformative censoring. When the 
		censoring depends on some latent variables, such as the risk, then the KM 
		estimator could be biased \citep{campigotto2014impact}. When censoring is 
		dependent on other variables, one can use imputation approaches for missing 
		data before estimating $S(t)$. Though we discussed at the beginning of 
		Section 2.2 that patients may stop visiting the hospital if their physical 
		conditions are too weak, such nonignorable censoring is usually close to the 
		event, because patients with rare diseases usually rely on visiting hospitals 
		regularly for examination or therapies. Based on numerical experiments, the 
		simple imputation strategy we provided in Section 2.2 helps the ZIKQ method 
		maintain its robust performance reasonably well. Extending the estimate of 
		$S(t)$ to incorporate other covariates, which may affect the censoring scheme 
		in complex scenarios, would also be a future interest. In practice, different types 
		of nonignorable missingness depending on the missed variables could happen 
		during data collection. Then, more sophisticated approaches are required for 
		model identifiability and the modeling of the missingness mechanism. 
		Approaches based on shadow variables are widely used to address model 
		identifiability while modeling missingness mechanisms through 
		parametric/semiparametric methods 
		\citep{shao2013estimation,miao2015identification,zhao2017reducing,zhao2018optimal,zhao2021versatile}.
		 It would be a future interest to explore different missing data mechanisms 
		under the current framework. Researchers are also encouraged to conduct 
		sensitivity analyses to evaluate the best and the worst cases if the independent 
		assumption is violated.

		\section*{Supplementary Materials}
		(1) The Supplement contains additional simulation results on bandwidth 
		selection and the proof for Theorem 1. (2) The related \texttt{R} code is 
		available at Github (\text{https://github.com/tianyingw/ZIKQ}). All tables and 
		figures can be reproduced based on 
		the descriptions in Section \ref{s:simulation}. 
		\par
		\section*{Acknowledgements}
		Y. W. thanks for support from the National Institutes of Health (HG008980 and 
		1RF1AG072272) and the National Science Foundation (DMS-1953527).
		\par
		

		\bibhang=1.7pc
		\bibsep=2pt
		\fontsize{9}{14pt plus.8pt minus .6pt}\selectfont
		\renewcommand\bibname{\large \bf References}
		\expandafter\ifx\csname
		natexlab\endcsname\relax\def\natexlab#1{#1}\fi
		\expandafter\ifx\csname url\endcsname\relax
		\def\url#1{\texttt{#1}}\fi
		\expandafter\ifx\csname urlprefix\endcsname\relax\def\urlprefix{URL}\fi

		\bibliographystyle{chicago}      
		\bibliography{refs}   

	\end{document}